\documentclass{aa} 

\usepackage[nolist]{acronym}

\usepackage{graphicx}
\usepackage{threeparttablex}
\usepackage{txfonts}
\usepackage{lastpage}
\usepackage{gensymb}

\usepackage{amsmath,amsfonts,amssymb}
\usepackage{bm}   
\usepackage{graphicx}
\usepackage[colorlinks=true, allcolors=blue]{hyperref}
\usepackage{astro_bib_macro}
\usepackage{ulem}

\usepackage{booktabs}

\usepackage{listings}

\graphicspath{{./}{./figures/}}   

\makeatletter
\renewcommand*\aa@pageof{, page \thepage{} of \pageref*{LastPage}}
\makeatother

\begin{document}

\title{Coherent differential imaging of high-contrast extended sources with VLT/SPHERE\thanks{Based on observations collected at the European Southern Observatory, Chile, 113.26G2}}

\titlerunning{CDI with VLT/SPHERE}
\authorrunning{Potier {et al.}}

\author{Axel Potier\inst{\ref{inst-a}} \and Raphaël Galicher\inst{\ref{inst-a}}\and Pierre Baudoz\inst{\ref{inst-b}} \and Johan Mazoyer\inst{\ref{inst-b}}\and Zahed Wahhaj\inst{\ref{inst-c}}\and Ruben Tandon\inst{\ref{inst-d}}\and Jonas G. Kühn\inst{\ref{inst-d}}\and Laura Perez\inst{\ref{inst-e}}\and Gael Chauvin\inst{\ref{inst-f},\ref{inst-g}}}

\institute{
LIRA, Université Paris Cité, Observatoire de Paris, Université PSL, Sorbonne Université, CNRS, F-92190 MEUDON, France\label{inst-a}
\and
LIRA, Observatoire de Paris, Université PSL, CNRS, Sorbonne Université, Université Paris Cité, 5 place Jules Janssen, 92195 Meudon, France\label{inst-b}
\and
European Southern Observatory, Alonso de Cordova 3107, Vitacura, Santiago, Chile\label{inst-c}
\and
Division of Space and Planetary Sciences, University of Bern, Sidlerstrasse 5, 3012 Bern, Switzerland\label{inst-d}
\and
Departamento de Astronomìa, Universidad de Chile, Camino El Observatorio 1515, Las Condes, Santiago, Chile\label{inst-e}
\and
MPIA, Max-Planck-Institut für Astronomie, Königstuhl 17, Heidelberg, D-69117, Germany\label{inst-f}
\and
Laboratoire J.-L. Lagrange, Université Cote d’Azur, CNRS, Observatoire de la Cote d’Azur, 06304 Nice, France\label{inst-g}\\
\email{axel.potier@obspm.fr}
}

\date{Received 28 July 2025; accepted 19 October 2025}

\abstract
  {High-contrast imaging relies on advanced coronagraphs and adaptive optics (AO) to attenuate the starlight. However, residual aberrations, especially non-common path aberrations between the AO channel and the coronagraph channel, limit the instrument performance. While post-processing techniques such as spectral or angular differential imaging (ADI) can partially address those issues, they suffer from self-subtraction and inefficiencies at small angular separations or when observations are conducted far from transit.}
  {
    We previously demonstrated the on-sky performance of coherent differential imaging (CDI), which offers a promising alternative. It allows for isolating coherent starlight residuals through speckle modulation, which can then be subtracted from the raw images during post-processing. This work aims to validate a CDI method on real science targets using VLT/SPHERE, demonstrating its effectiveness in imaging almost face-on circumstellar disks, which are typically challenging to retrieve with ADI.
  }
  {
    We temporally modulated the speckle field in VLT/SPHERE images, applying small phase offsets on the AO deformable mirror while observing stars surrounded by circumstellar material: HR~4796A, CPD-36~6759, HD~169142, and HD~163296. We hence separated the astrophysical scene from the stellar speckle field, whose lights are mutually incoherent. 
  }
  {
    Combining a dozen of data frames and reference coronagraph point spread functions through a Karhunen–Loève image projection framework, we recover the circumstellar disks without the artifacts that are usually introduced by common post-processing algorithms (e.g., self-subtraction).
  }
  {The CDI method therefore represents a promising strategy for calibrating the effect of static and quasi-static aberrations in future direct imaging surveys. Indeed, it is efficient, does not require frequent telescope slewing, and does not introduce image artifacts to first order.}
   

   \maketitle

\section{Introduction}
\label{sec:introduction}

High-contrast imaging aims at the detection of light emitted, reflected, or scattered by any object (exoplanet, disk) in circumstellar environments. It potentially allows for a precise spectral, astrometric, and polarimetric characterization of any detected object. Such observations require cutting-edge coronagraph technologies and state-of-the-art wavefront sensing and control (WS\&C) capabilities~\citep{Galicher2023} that are combined in dedicated facilities such as VLT/SPHERE \citep{Beuzit2019}, Gemini/GPI \citep{Macintosh2015a}, Subaru/SCExAO \citep{Lozi2018}, and Magellan/MagAO-X \citep{Males2020}. However, the performance of these instruments remains limited by residual stellar light reaching the science detector, thereby drowning the exoplanet signal. These residuals are mainly caused by the spatiotemporal limitations of the extreme adaptive optics (XAO) system, creating a smooth halo topped with stochastic stellar speckles in the long-exposure coronagraph image. The topping speckles, which result from unaveraged AO residuals, vary from one exposure to the other. The longer the exposure, the fainter the speckles~\citep{Singh2019}. Additionally, the XAO wavefront sensor (WFS) is insensitive to the optical errors located in the science channel after the beam splitter and is biased by the aberrations in the WFS channel, leaving residual aberrations at the coronagraph. These non-common path aberrations \citep[NCPA,][]{Fusco2006} subsequently introduce additional stellar leakage in the form of static or quasi-static speckles. It is also worth noting that amplitude aberrations caused by beam transmission inhomogeneities or out-of-pupil phase errors are not typically corrected by conventional XAO WS\&C algorithms. These amplitude aberrations also induce quasi-static stellar speckles in the coronagraphic image.

The active correction of these quasi-static speckles is a dynamic field of research. Techniques based on a focal plane WFS are being tested or implemented on ground-based instruments. Some of these methods are implemented on a daily basis, leveraging the calibration source to estimate and correct for the NCPAs \citep{Sauvage2007,NDiaye2016b,Lamb2018}. However, once on-sky, these pre-calibrations are limited by differential aberrations between the internal source unit and the science path \citep{Vigan2019,Potier2022c} as well as by a temporal decorrelation of the NCPAs \citep{Milli2016,Vigan2022}. Therefore, the on-sky calibration (directly using the photons from the science target star) of the NCPAs was tested with various instruments \citep[e.g.,][]{Martinache2014, Galicher2019, Vigan2019, Bos2021, Skaf2022, Xin2023, Kueny2024}. In particular, a dark hole demonstration -- where destructive interferences are set up on a region of a detector -- has been performed in narrowband \citep{Potier2022b} and broadband light \citep{Galicher2024}. However, in poor observing conditions (meaning unstable AO correction), the dark hole technique becomes inefficient, as wavefront sensing must be repeated several times at each iteration, requiring a stable AO correction over several minutes.

Because of these drawbacks, the current strategy to mitigate speckles relies on post-processing approaches, each relying on such particular observational techniques as angular \citep[ADI,][]{Marois2006}, spectral \citep[SDI,][]{Racine1999}, polarimetric \citep[PDI,][]{Kuhn2001}, and reference-star \citep[RDI,][]{Lafreniere2009} differential imaging. Building on such classical strategies, algorithms such as the Locally Optimized Combination of Images (LOCI, \cite{Lafreniere2007}) and the principal component analysis (PCA, \cite{Amara2012}), also known as Karhunen-Loève Image Processing (KLIP, \cite{Soummer2012}), have been developed. Additionally, numerous optimizations have been introduced, such as adjusting the size of the reduction zones \citep{Marois2014, Ren2023} or modifying the selection of linear components \citep{Gomez2016}. Also, alternative approaches such as the patch covariance algorithm (PACO, \cite{Flasseur2018}) have been developed, which leverage prior knowledge of speckle noise distribution to better distinguish planetary signals. However, these techniques suffer from the self-subtraction of off-axis sources, especially at small angular separations \citep{Esposito2014}.
Spectral-differential imaging self-subtraction depends on the separation and on the spectral properties of the object \citep{Rameau2015} but can be mitigated through higher spectral bandwidth \citep{Gerard2019b}. Angular-differential imaging self-subtraction can be quantified and calibrated based on the evolution of the parallactic angle (PA) and the specific parameters used in the data reduction process. But, face-on disks remain undetected through ADI because of the centrosymmetric geometry of such systems. To address this limitation, forward modeling methods have been developed for point source objects \citep{Marois2010b, Pueyo2016} or disk analysis \citep{Mazoyer2020}. However, in the case of disks, a wide range of possible morphologies must be explored, which is computationally intensive. Moreover, the diversity of complex and irregular structures cannot be fully captured by existing models.

More recently, coherent differential imaging (CDI) techniques -- which should not suffer from these limitations -- have been developed, leveraging the advent of focal plane WS\&C algorithms cited earlier \citep{Bottom2017, Gerard2018b, Potier2022b}. These methods rely on the coherent modulation (either spatial or temporal) of the speckle field. Only the coherent signal (starlight) responds to the modulation and, therefore, can be isolated from the astrophysical signal in post-processing. Contrary to \cite{Bottom2017}, our previous CDI demonstration using the VLT/SPHERE High Order Deformable Mirror (HODM) to sequentially modulate the speckle field \citep{Potier2022b} was not validated on a science target. In the current paper, we validate the CDI algorithm, observing four circumstellar disks with VLT/SPHERE. In Sec.~\ref{sec:method} we describe the acquisition of CDI sequences and their combination to isolate the disk signals in post-processing. In Sec.~\ref{sec:results} we present the resulting images and derive intensity measurements to analyze the performance of our methods. Finally, we discuss the advantages and limitations of the algorithm as well as potential improvements in Sec.~\ref{sec:discussion}.

\section{Method}
\label{sec:method}

\subsection{Single CDI sequence workflow}
\label{subsec:CDI_algorithm}
\begin{figure*}
    \centering
    \includegraphics[width=\linewidth]{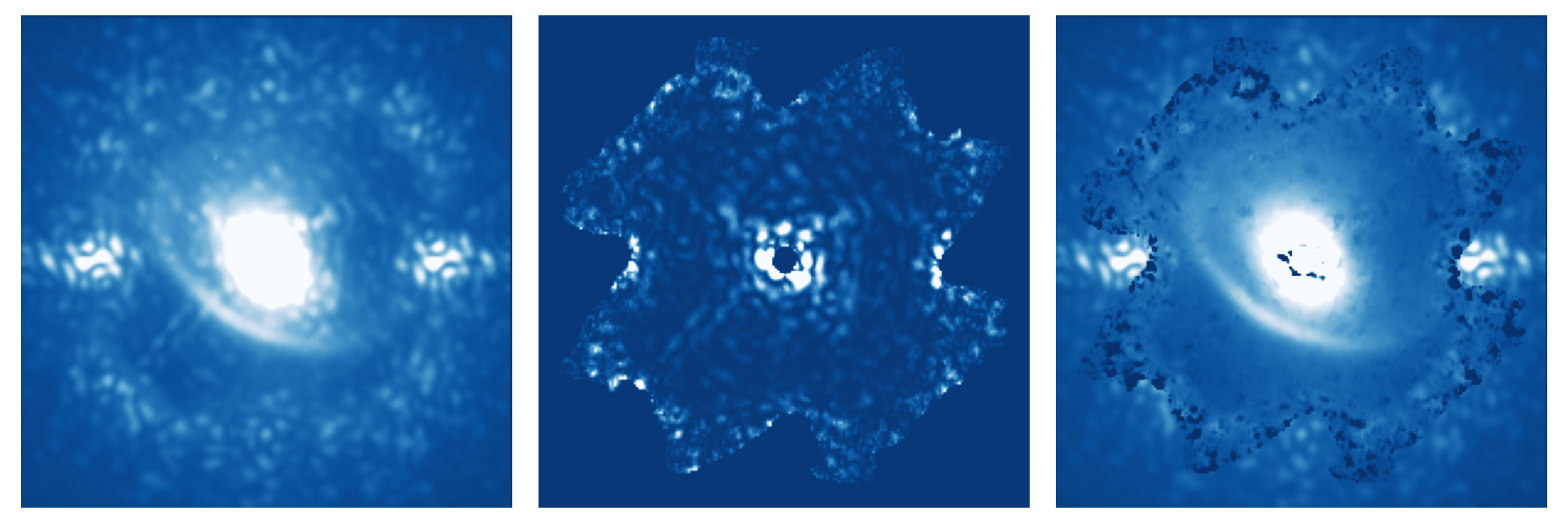}
    \caption{Image decomposition of HD~163296 through CDI. Left: Coronagraph image with no probe (i.e., total intensity image). Center: PWP-estimated speckle field. Right: CDI result (estimated astrophysical scene subtracted from the total intensity image) after one sequence of PWP. The negative values, indicated by black-colored regions, show local over-subtraction in areas where PWP cannot properly estimate the speckle field.}
    \label{fig:1iter_example}
\end{figure*}
The CDI method based on pair-wise probing (PWP) has been intensively described in the literature \citep[see, e.g.,][]{GiveOn2007, Potier2020a, Potier2020b}, and its equations are detailed in Appendix~\ref{sec:PWP_algorithm}. It relies on a temporal modulation of the speckle field using the HODM. The diversities (or probes) are applied to the HODM by modifying the WFS reference slopes with the XAO system running in closed-loop. Contrary to the study performed in \cite{Potier2022b}, three pairs of probes are used to cover the full 360\degree~field of view. For the reasons described in \cite{Laginja2025} (i.e., reduction of mechanical constraints and nonlinear coupling, knowledge of the HODM influence functions, and mitigation of algorithm nonlinearities), we consecutively poked a triplet of neighboring single-actuators distributed along both the HODM axes, with a peak-to-valley amplitude of 400~nm. Such an amplitude was chosen to obtain a sufficient signal-to-noise ratio on the probe images while remaining in the linear regime defined in Appendix~\ref{sec:PWP_algorithm}. For each applied diversity, one image was acquired on the science detector IRDIS \citep{Dohlen2008}; hence, six images were required to estimate the modulated signal. Their combined exposure time was arbitrarily chosen to be equal to the exposure time of the raw signal with no probe applied. Therefore, a CDI sequence took twice as long as a typical data acquisition process. We propose some workarounds in Sec.~\ref{sec:discussion}.

We first estimated the speckle electric field (E-field) through the PWP algorithm, as described in Appendix~\ref{sec:PWP_algorithm}. The square modulus of the field, called $I_{PWP,i}$, was then subtracted from the total intensity image (obtained with no probe) $I_{tot,i}$ of the $i$th CDI sequence. Fig.~\ref{fig:1iter_example} represents an illustration of one CDI sequence. The raw total intensity image consists of: 1)~a smooth halo of XAO residuals with a coherence time below 10~ms (see Appendix \ref{sec:observing_sequences}), overlaid with topping speckles; 2)~individual static speckles induced by NCPAs, amplitude aberrations, and the diffraction pattern of the coronagraph; and 3)~the astrophysical signal to be extracted. Because the CDI algorithm operates with seconds-long probe images to average out the intensity variations caused by atmospheric turbulence \citep{Singh2019} (and achieve a sufficient S/N), the XAO halo becomes incoherent and cannot be sensed by PWP. Static speckles alone ($I_{PWP,i}$) are therefore estimated in the center image of Fig.~\ref{fig:1iter_example}. Subsequently, the resulting image after one CDI sequence ($I_{CDI}$) contains the smooth XAO halo and the astrophysical signal with no self-subtraction.

\subsection{Combination of CDI sequences}
\label{subsec:frame_combination}
The CDI algorithm can be performed in parallel to the dark-hole algorithm, but only on an individual sequence, since the speckle intensity distribution varies at each iteration \citep{Potier2022b}. However, standalone CDI (with a static speckle pattern, which is not the case during a dark hole convergence loop) enables the combination of reference images by relying on previous methods that improve reference differential imaging algorithms \citep{Lafreniere2007, Soummer2012}.

The method can first be generalized using the following mathematical notations. For $N$ CDI sequences, the reference image $I_{ref,i}$ -- containing the static speckle signal -- is reconstructed and then subtracted from $I_{tot,i}$ in order to retrieve the astrophysical signal ($I_{CDI,i}$). To ensure the speckles remain quasi-static during observations, the command of the instrument derotator is set such that the pupil does not rotate. Pupil tracking mode also guarantees that the calibrated model of the instrument, represented by the coronagraph operator $C$ in Appendix~\ref{sec:PWP_algorithm}, remains consistent along all the sequences. Then, for the sake of increasing the S/N, $I_{CDI,i}$ is derotated by an angle $\theta_i$ corresponding to the PA at sequence $i$. The result is then averaged (the median can also be used) over the full sequence,
\begin{equation}
\label{eq:frame_combination}
    I_{CDI} = \frac{1}{N}\sum_{i=1}^NR_{\theta_i}[I_{tot,i} - I_{ref,i}],
\end{equation}
where $R$ corresponds to the rotation matrix, while $I_{CDI}$ is the result of the algorithm. The main objective of the method is to determine $I_{ref,i}$ as accurately as possible with respect to the static stellar residuals, using the sequence of $I_{PWP,i}$.

\subsubsection{Batch process}
\begin{figure}
    \centering
    \includegraphics[width=\linewidth]{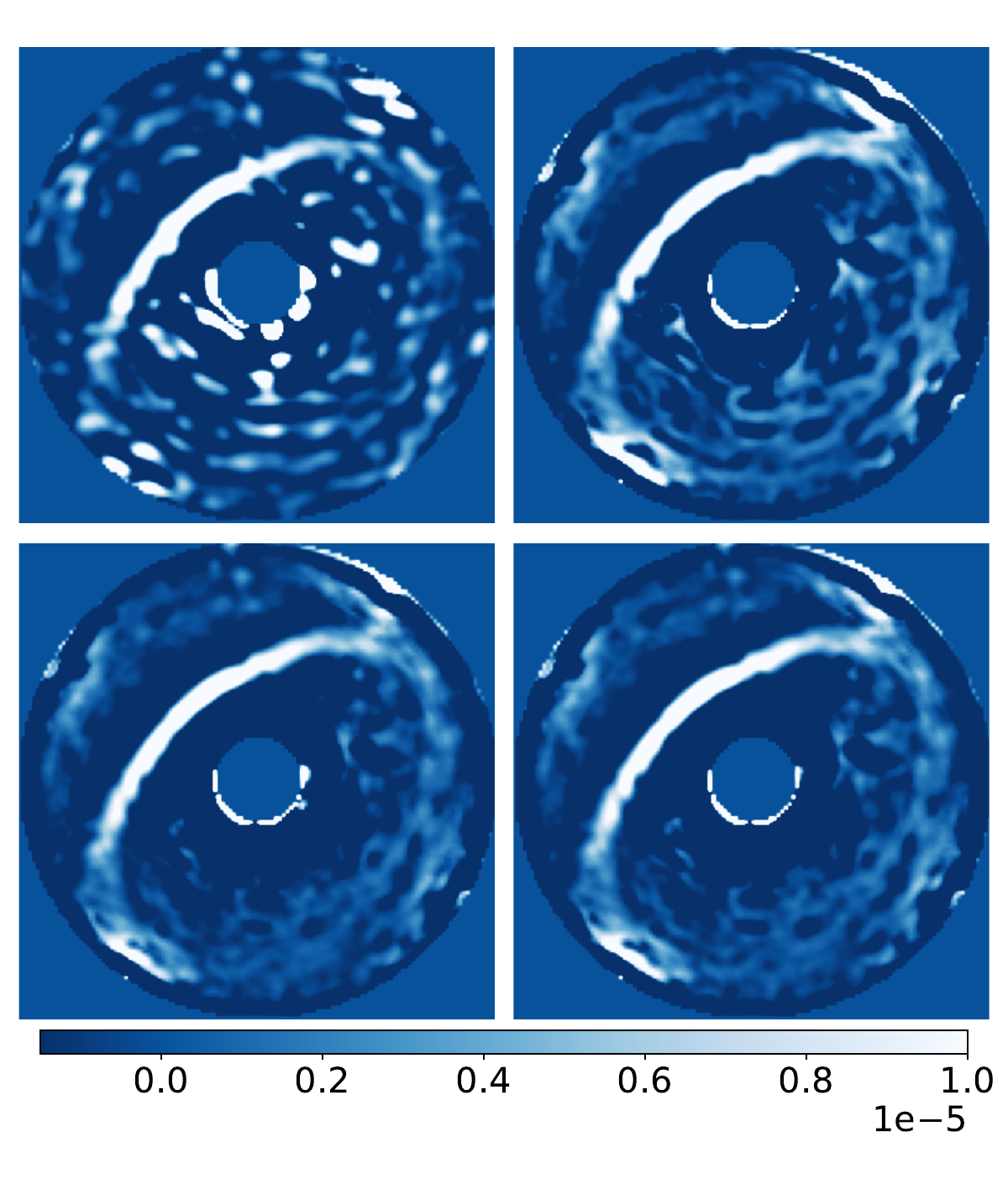}
    \caption{$I_{CDI}$ results for different reference estimations. Top left: $I_{ref,i}=0$, no CDI post-processing applied (equivalent to noADI). Top right: $I_{ref,i} = I_{PWP,i}$, a simple batch process used. Bottom left: Karhunen-Loève coherent component projection. Bottom right: Karhunen-Loève image projection. All images have been high-passed filtered, as explained in Sec.~\ref{subsec:highpass_filter}.}
    \label{fig:Comparison_algorithms}
\end{figure}

Since PWP aims to retrieve the stellar speckle E-field alone, the most straightforward algorithm is a simple batch process, where the reference image at each sequence is simply the squared modulus of the PWP E-field estimation for the same sequence:
\begin{equation}
    I_{ref,i} = I_{PWP,i}.
\end{equation}
To illustrate its limitations, the final result of this option is shown in Fig.~\ref{fig:Comparison_algorithms} (top right) and compared to more advanced approaches (see Sec.~\ref{subsubsec:PCA_coherent} and ~\ref{subsubsec:khi-squared}). Although the image quality is improved with respect to the noADI case ($I_{ref,i} =$), speckle residuals are not perfectly canceled. We suspect that the inaccuracy of $I_{PWP,i}$ is due to PWP model uncertainties and nonlinearities, photon noise, as well as strong variations in atmospheric turbulence -- not in a steady state -- that bias the recording of probe images (see, e.g., the bright estimated speckles at small angular separations that over-subtract the total intensity image in Fig.~\ref{fig:1iter_example}). As for ADI or RDI, more advanced sequence combinations can therefore be key to mitigating the latter two noise components in the reference image. The number of required CDI sequences is discussed in Sec.~\ref{subsec:results_vs_frame}.

\subsubsection{Karhunen–Loève coherent component projection}
\label{subsubsec:PCA_coherent}
\begin{figure}
    \centering
    \includegraphics[width=\linewidth]{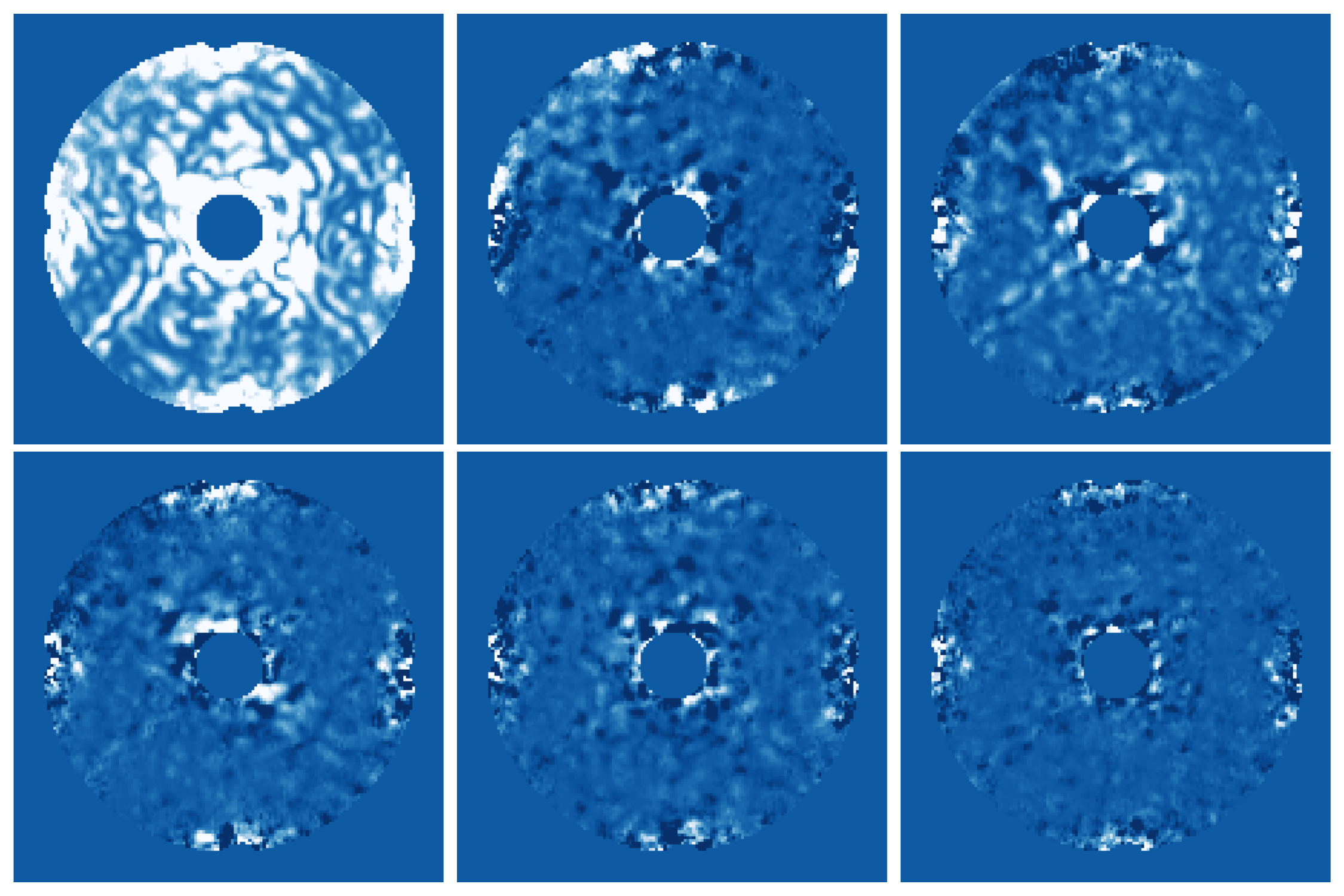}
    \caption{Illustration of the $M_{PWP}$ eigenvector decomposition. Top row: Principal component 1, 2, and 3. Bottom row: Principal component 5, 10, and 15.}
    \label{fig:Eigenvectors}
\end{figure}

We now assume a library of reference coherence intensity components obtained with PWP, whose singular value decomposition is given by
\begin{equation}
\label{eq:SVD}
    M_{PWP} = [I_{PWP,1}, I_{PWP,2}, ..., I_{PWP,N}] =  U\Sigma V^T,
\end{equation}
where $U$ and $V$ are two square matrices with orthogonal vectors, called the left and right singular vectors, respectively, and $\Sigma$ is a rectangular diagonal matrix containing the singular values~$\sigma_i$ of~$M_{PWP}$, sorted in ascending order. In this framework, the principal components or eigenvectors of~$M$, described in the pixel space, are the vectors of $V^T$. They are shown in Fig.~\ref{fig:Eigenvectors} for the observation of HD~163296 (16~PWP sequences, see Tab.~\ref{tab:data_acquisition}). As expected the first eigenvectors exhibit a speckle-shaped intensity distribution, while the last eigenvectors display a noisier distribution at the pixel scale. It means that truncating the highest $N-K_{klip}$ eigenvector components in the $I_{PWP,i}$ preserves the speckle intensity distribution displayed in most of the PWP estimates while reducing high-frequency noise.
Mathematically, using the notations from \cite{Soummer2012}, the reference images are the projection of the PWP estimate onto the truncated PWP eigenvector basis:
\begin{equation}
    I_{ref,i}(K_{klip}) = \sum_{k=1}^{K_{klip}}<I_{PWP,i},V^T_k>V^T_k = (\left.U\Sigma_{K_{klip}}V^T)\right|_{i},
\end{equation}
where
\begin{equation}
    \Sigma_{K_{klip},ij} = \left\{
    \begin{array}{ll}
        \Sigma_{ij} & \mbox{if } i<K_{klip} \\
        0 & \mbox{otherwise.}
    \end{array}
\right. 
\end{equation}
The $K_{klip}$ threshold value is then optimized with respect to the current dataset and observing conditions. We note that such an analysis can also be directly performed on a library of probe image differences.

The former projection aims at tracking and calibrating speckle intensity modes that vary at minute timescales along the CDI sequences. However, if pixel-scale noise is significant, the solutions can simply be averaged over all PWP sequences to be used as the image reference:
\begin{equation}
    I_{ref,i} = \frac{1}{N}\sum_{i=1}^{N} (\left.U\Sigma_{K_{klip}}V^T)\right|_{i}.
\end{equation}
Such a solution is more robust to noise, but it only targets perfectly static speckles throughout the sequences.

\subsubsection{Karhunen–Loève image projection}
\label{subsubsec:khi-squared}
Instead of projecting the eigenvectors computed in Eq.~\ref{eq:SVD} onto the coherent signal, the standard approach is the Karhunen–Loève image projection \citep[KLIP,][]{Soummer2012}, where the eigenvectors are directly projected onto each total intensity image:
\begin{equation}
    I_{ref,i}(K_{klip}) = \sum_{k=1}^{K_{klip}}<I_{tot,i},V^T_k>_SV^T_k,
    \label{eq:cdi_seq_klip}
\end{equation}
where $S$ corresponds to the search area. However, since this method aims to project static speckle components onto images that also contain the astrophysical source and a bright AO halo, it may result in an over-subtraction of the static speckle in order to match the halo brightness. Hence, $S$ may be chosen in areas where the intensity of the static speckles dominates, i.e., where the turbulence halo is the faintest typically in the outer part of the HODM influence function. In this paper, we chose $S$ to be an annulus, ranging from 551 to 919~mas for all the disks. On the other hand, KLIP is convenient because it enables the use of such widely employed advanced tools and software as VIP \citep{Gomez2017,Christiaens2023}.

\subsection{High-pass filter}
\label{subsec:highpass_filter}
We explained why the technique can neither target the XAO smooth halo nor its topping speckles. As mentioned the topping speckle intensity is minimized while recording long exposures. The XAO halo remains in the post-processed images. To extract the astrophysical signal, this halo could potentially be estimated through AO telemetry and then filtered in the future. Instead, assuming the halo to be smoother than the astrophysical signal, the processed image $I_{CDI}$ can be high-pass filtered, for instance using a Gaussian kernel,
\begin{equation}
    I_f = \mathcal{G}[I_{CDI}(K_{klip}),\sigma_{hp}]
\label{eq:high_pass},
\end{equation}
where $\mathcal{G}$ represents the high-pass Gaussian filter operator and $\sigma_{hp}$ is the standard deviation of the Gaussian kernel. Contrary to the previous steps of the reduction, this last procedure causes the subtraction of the astrophysical signal, especially for large structures, resulting in a reduced throughput. $\sigma_{hp}$ therefore must be chosen wisely, depending on the structures one aims to image. Sec.~\ref{subsec:highpass_filter_effect} discusses its repercussions more quantitatively for our dataset. It is worth noting that alternative CDI methods using faster science detectors may be able to calibrate fast speckles induced by atmospheric turbulence \citep{Kuhn2018b, Gerard2019, Thompson2022}.

\section{Results of disk imaging}
\label{sec:results}
\begin{figure*}[h!]
    \centering
    \includegraphics[width=0.615\linewidth]{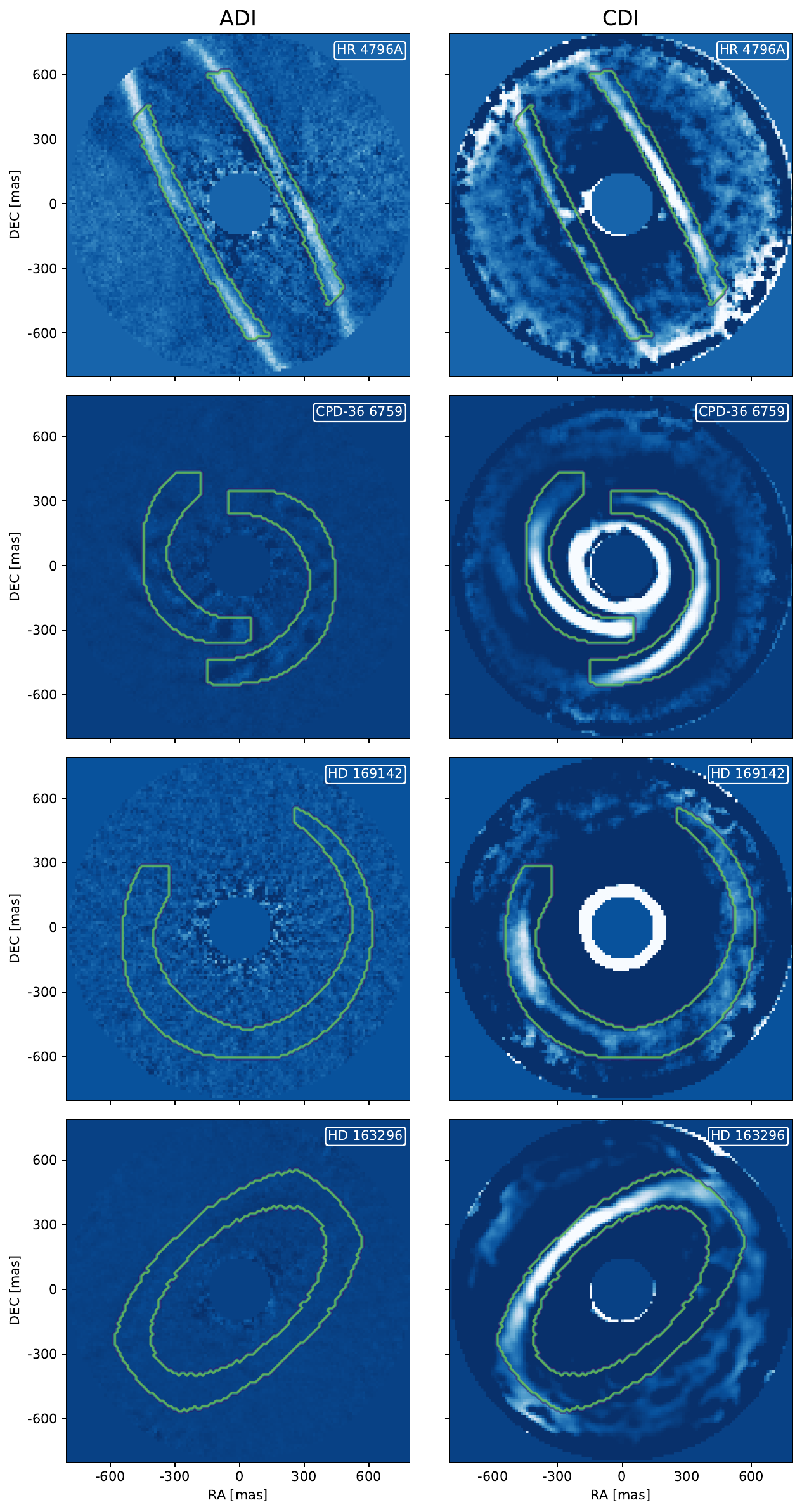}
    \includegraphics[width=0.33\linewidth]{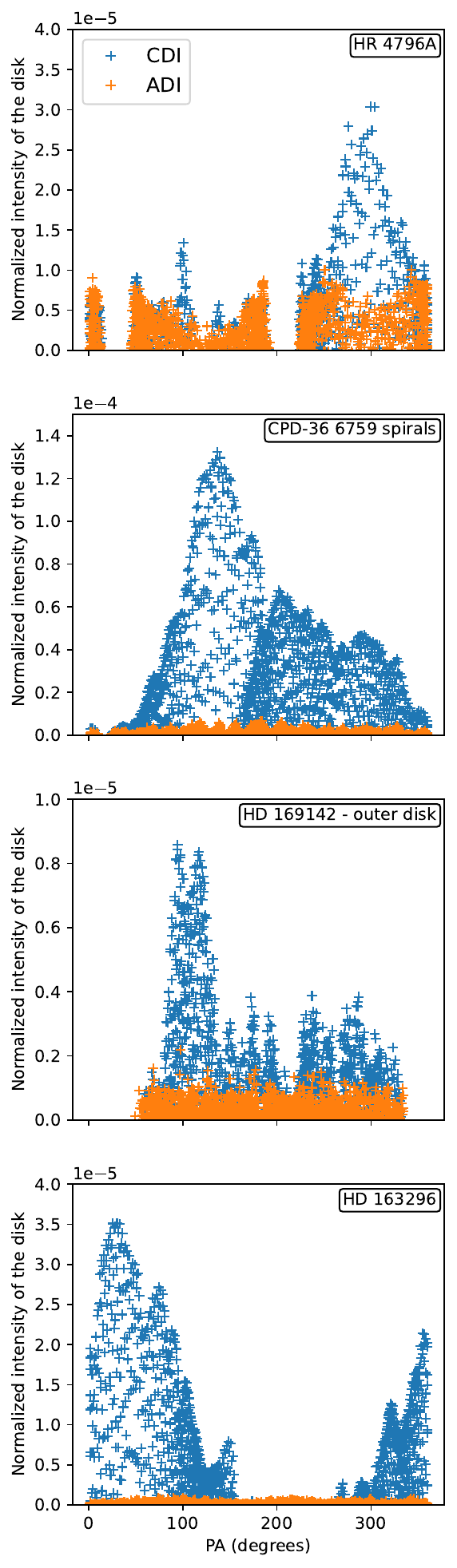}
    \caption{Comparison of disk images obtained with ADI (left column) and CDI (center column). For each disk, the ADI and CDI images are displayed using the same color scale; however, the scales vary from one disk to another. ADI and CDI results were obtained through the same dataset. We use K-klip ADI \citep{Soummer2012} with $K_{klip} = N_{filt}/2$. Right column: Comparison of disk intensities, for pixels inside the region encircled in green, obtained with both ADI and CDI as a function of the position angle.}
    \label{fig:Images_disk}
\end{figure*}
Coherent differential imaging was performed on VLT/SPHERE on June 4, 2024. The observing conditions are described in Appendix~\ref{sec:observing_sequences}. Four targets with known circumstellar and almost face-on disks were observed throughout the night. Their respective data acquisition is detailed in Tab.~\ref{tab:data_acquisition}. In addition to the set of $N$-acquired CDI sequences, $N_{filt}$ CDI sequences were down-selected by eye for post-processing, getting rid of the few sequences where bursts of seeing occurred and led to either a low-quality total intensity image or a misestimate of the coronagraph point spread function (PSF) through PWP.
\begin{table}
    \caption{Observations from 2024-06-04}
    \centering
    \begin{tabular}{ccccccc}
    \hline
    \hline
        System & Obs. date & $t_{\text{exp}}$ & $t_{\text{p}}$ & $N$ & $N_{filt}$ & $\Delta \text{PA}$\\
         & (UTC) & (s) & (s) &  &  & (degree)\\
        \hline
        HR~4796A & 23:46:13 & 96 & 16 & 16 & 9 & 57.7 \\
        CPD-36~6759 & 01:12:46 & 96 & 16 & 20 & 15 & 52.5 \\
        HD~169142 & 03:36:10 & 192 & 32 & 16 & 11 & 30.4 \\
        HD~163296 & 06:04:23 & 96 & 16 & 16 & 14 & 14.7\\
    \end{tabular}
    \tablefoot{The columns give the host star name, the time the observation began, the exposure time~$t_{exp}$ of the no-probe coronagraphic images, the exposure time~$t_p$ of the probe images, the number~$N$ of CDI sequences, the $N_{filt}$ of selected CDI sequences for data processing, and $\Delta \text{PA}$, the total change of PA throughout the observing sequences.}
    \label{tab:data_acquisition}
\end{table}

The final post-processed images obtained with~Eqs.~\ref{eq:cdi_seq_klip}~and~\ref{eq:high_pass} are shown in Fig.~\ref{fig:Images_disk} (central column) together with the ADI results obtained through the same dataset, discarding probe images for comparison (left column).
On one hand, in addition to the speckle intensity being minimized with respect to noADI (see in Fig.~\ref{fig:Comparison_algorithms}), CDI results show that the disk features are well recovered by the algorithm, while self-subtraction is absent to first order -- even for face-on disks. On the other hand, because ADI tends to subtract the signal of interest \citep{Milli2012, Ruane2019b, Xie2022}, almost no signal is detected for three of our targets using ADI, and the HR\,4796 disk image is impacted by self-subtraction particularly at small angular separations.

To quantitatively compare the performance of CDI with respect to ADI in retrieving disk structures, we measured the normalized intensity of each pixel within the green area overplotted in Fig.~\ref{fig:Images_disk}, which corresponds to the estimated disk locations. These intensities were then plotted as a function of their respective position angles in Fig.~\ref{fig:Images_disk} (right column). The results show that in all cases, ADI suffers from self-subtraction — significantly reducing the disk signal — although CDI remains largely unaffected. For example, the mean intensity ratio between CDI and ADI across these pixels reaches 216 for HD~163296. In the case of HR~4796A, although the average ratio is lower (approximately 7) due to the disk’s inclination, which makes it less vulnerable to ADI self-subtraction, and poorer observing conditions, values as high as $\sim$100 are observed for pixels closest to the disk center.

Regardless, all these results demonstrate that CDI is a promising alternative to ADI, mostly free of any self-subtraction artifacts, especially for spatially extended sources such as disks. Coherent differential imaging is also a serious contender to RDI, as it requires neither extra telescope slewing and exposure time on a nonscience target \citep{Wahhaj2021}, nor the need to build up an archival PSF library \citep{Ruane2019b, Xie2022, Ren2023}. However, without further observations of targets devoid of extended structures, deriving reliable and general contrast curve comparisons with other post-processing methods is difficult. As a result, it remains uncertain whether CDI could serve as a viable replacement for ADI in point-source detection. This question will be addressed in future follow-up studies. Also, a deep comparison with alternative post-processing techniques such as RDI \citep[e.g.,][]{Xie2022}, PDI \citep[e.g.,][]{Gratton2019, Ren2023}, or advanced ADI pipelines designed to reduce self-subtraction of face-on disks (e.g., \cite{Juillard2024, Flasseur2024}), as well as the detailed analysis of disk structures or single sources in these images is beyond the scope of this paper.

\section{Discussion}
\label{sec:discussion}

\subsection{Influence of the high-pass filter}
\label{subsec:highpass_filter_effect}
\begin{figure}[t]
    \centering
    \includegraphics[width=\linewidth]{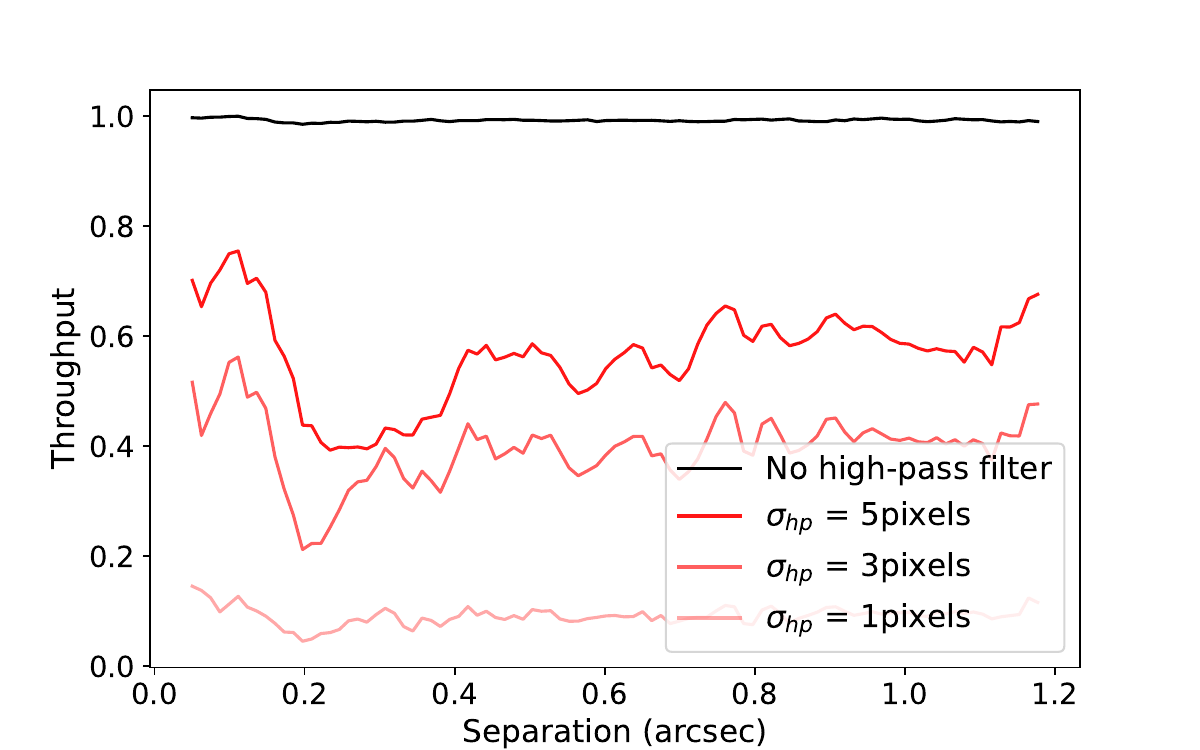}
    \caption{Effect of the high-pass filter on the off-axis source throughput for various Gaussian kernel standard deviations.}
    \label{fig:hpfilter_throughput}
\end{figure}

The main performance limitation of the current algorithm is its inability to tackle speckles whose lifetimes are smaller than the total exposure time of one CDI sequence. In particular, the dynamic speckles induced by the atmospheric turbulence are averaged into a smooth halo in long exposure images and filtered with a high-pass filter (see in Sec.~\ref{subsec:highpass_filter}). However, this final process also filters out the astrophysical signal that was unaffected until this stage. Fig.~\ref{fig:hpfilter_throughput} shows the influence of the high-pass filter on the entire algorithm transmission -- calculated with VIP -- for the HR~4796A disk-free dataset. Indeed, to prevent the disk signal from interfering with the contrast measurement, the disk was forward-modeled and subtracted from the dataset, according to the disk parameters detailed in \cite{Milli2017} and as described in Appendix~\ref{sec:FM_HR4796A}. While the planet transmission is scarcely impacted at any separation without high-pass filtering -- indicating the absence of self-subtraction with CDI -- the potential planet throughput decreases with smaller $\sigma_{hp}$. For instance, it decreases by $\sim60$\% for $\sigma_{hp}=3$~pixels, which is close to the PSF size. Other approaches to estimate and subtract this halo -- such as using WFS telemetry -- would therefore be essential to preserve the main benefit of CDI. SPHERE/IRDIS has a minimal exposure time of $\sim1$s, but other CDI strategies using faster science detectors have been proposed to enable quicker estimation of the speckle field, targeting dynamic speckles caused by atmospheric turbulence \citep{Kuhn2018b, Gerard2019}.

\subsection{Result versus number of CDI sequences}
\label{subsec:results_vs_frame}
\begin{figure}
    \centering
    \includegraphics[width=1\linewidth]{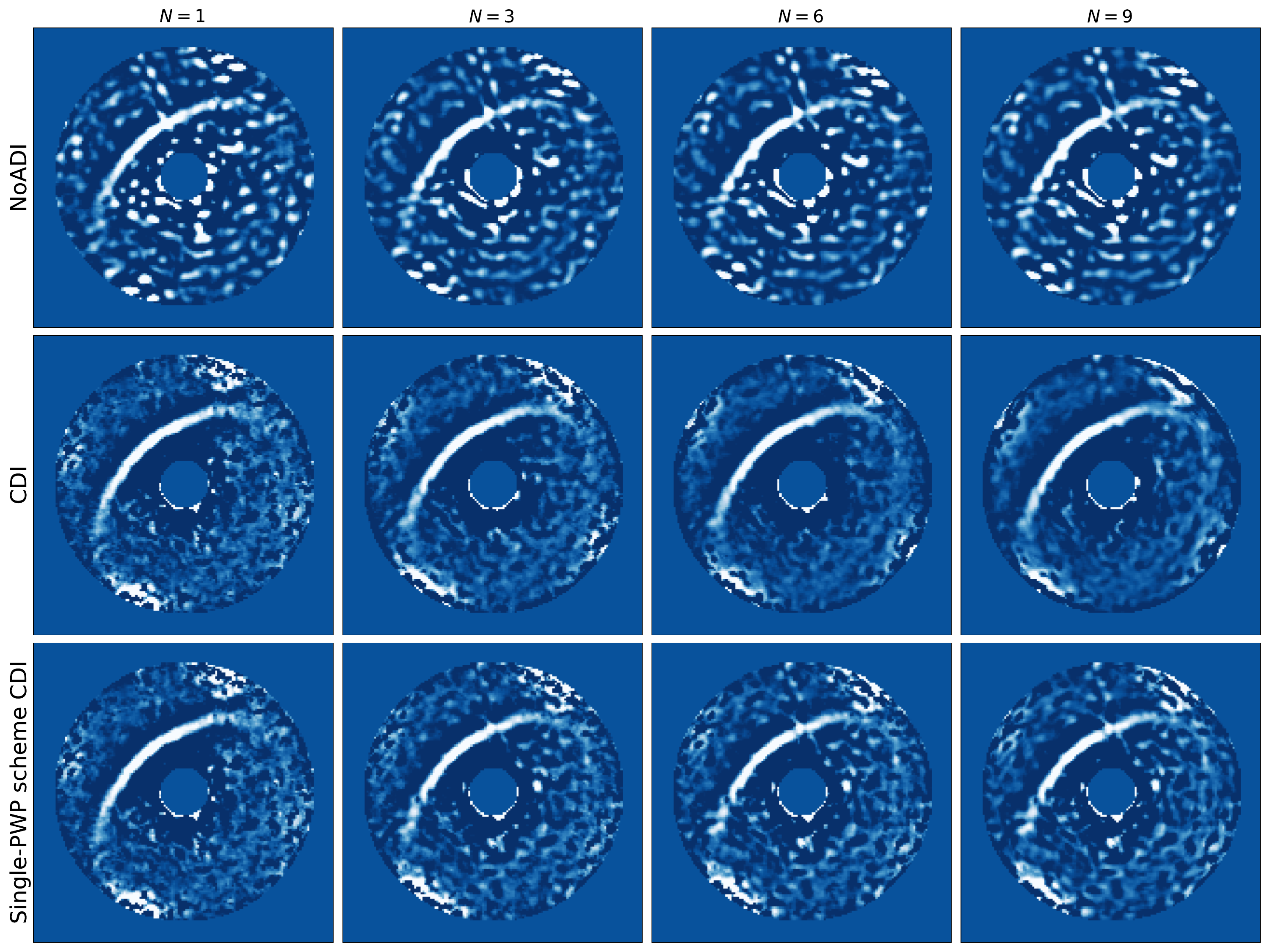}
    \caption{NoADI (top row), CDI through KL image projection (center row), and single-PWP scheme CDI (bottom row) final images with respect to the number of considered frames: one, three, six, and nine sequences.}
    \label{fig:Result_vs_framenb}
\end{figure}

It can be argued that the acquisition of the PWP library is a time-intensive process to the detriment of astrophysical signal acquisition. We show in Fig.~\ref{fig:Result_vs_framenb} the final combined-CDI images obtained throughout the $N$ CDI sequences, which are compared with the noADI case ($I_{ref,i}=0$ in eq.~\ref{eq:frame_combination}, \cite{Galicher2018}). It demonstrates that CDI leads to better performance for any number of sequences, particularly during the first few, when the static speckles do not average with diurnal motion. This means the CDI strategy can be quickly applied on targets of interest -- much faster than ADI, which requires the planet image to rotate in the field of view (FoV), especially at low angular separations -- such as during surveys when many systems must be explored within a limited time. 

Additionally, only a few CDI schemes were acquired at the beginning of the observation sequence, followed by more standard acquisitions, assuming the speckle field remains static over time. This option, called single-PWP scheme CDI, is also illustrated in Fig.~\ref{fig:Result_vs_framenb}, where it is assumed that only one PWP intensity estimate is performed at the beginning of the sequence to retrieve the astrophysical scene (i.e., $I_{ref,i}=I_{PWP,1}$ for all $i$). First, we see that this high-speed speckle calibration constantly outperforms the noADI case. Also, the final image after one sequence is
identical to the KL-CDI result. Although the speckle subtraction performance degrades afterward, it saves $6N$ probe acquisitions, which in our case can be utilized to double the exposure time dedicated to image acquisition.

\subsection{CDI self-calibration}
\label{subsec:phase_closure}
\begin{figure}
    \centering
    \includegraphics[width=\linewidth]{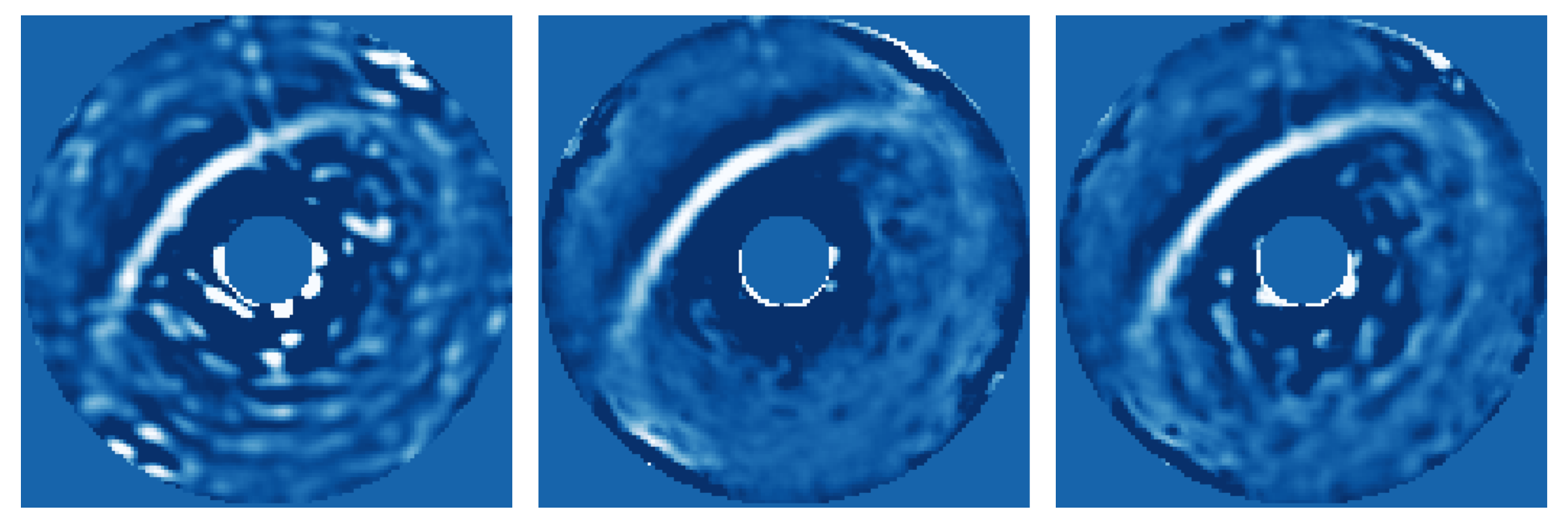}
    \caption{Comparison of signal extraction methods: the signal obtained from probed images (right, i.e. self-calibration using eq.~\ref{eq:selfcalibrationderivation}) versus conventional CDI (center, using eq.~\ref{eq:cdi_seq_klip}) and noADI (left). All results are high-pass filtered, as explained in Sec.~\ref{subsec:highpass_filter}.}
    \label{fig:Self_calibration}
\end{figure}

In addition to the former point, CDI sequences can self-calibrate, with the astrophysical signal extracted from the probe images themselves, similarly to phase closure in interferometry. In brief, after the speckle E-field $E_S$ is estimated from eq.~\ref{eq:PWP_matmul}, it can be added to or subtracted from the modeled probe E-field $iC[A\psi_m]$ to calibrate both the positive and negative probe images without the astrophysical signal $\Big|E_S\pm iC[A\psi_m]\Big|^2$. Subsequently from Eq.~\ref{eq:image_probe}, the astrophysical image $I_a$ can be retrieved via
\begin{equation}
I_a = I_{m\pm} - \Big|E_S\pm iC[A\psi_m]\Big|^2,
\end{equation}
where $I_{m\pm}$ are the probe images. The full mathematical derivation proposed in this paper to combine many CDI sequences is described in Appendix~\ref{sec:selfcalibrationderivation}. Such a strategy is demonstrated in Fig.~\ref{fig:Self_calibration}, where the probe images were self-calibrated to extract the HD~163296 disk. We, however, observe an additional amount of speckle noise in the final image caused by misestimated speckles that rotate in the FoV. This can be explained either by a model error occurring in $iC[A\psi_m]$ and/or by additional shot-noise error contained in the probe images. Nevertheless, CDI self-calibration can be a workaround to compensate for its slowness. It could also be used in parallel to dark hole algorithms to maximize science return in the context of space-based coronagraphs such as RST/CGI.

\subsection{Current PWP downsides}
\label{subsec:current_downsides}
\subsubsection{Nonlinearities}
Reducing the duty cycle is not the only limitation of the proposed strategy. Coherent differential imaging is also altered by several other drawbacks. First, PWP is inherently limited by nonlinearities arising when the probe amplitude is too high \citep{Groff2015}. Here, we chose high-amplitude probes, such as maximizing the S/N in the probe images, while the dark hole (DH) correction performed in \cite{Potier2022b, Galicher2024} still converged after a few iterations. However, in the case of standalone CDI, the post-processing might directly be biased by the nonlinear factor, limiting its contrast gain. Subsequent work will aim to quantitatively assess the best choice of probe amplitude with respect to S/N while preserving a high duty cycle. Nevertheless, previous studies have shown that single-actuator probes are less affected by nonlinearities than other shapes \citep{Laginja2025}.

\subsubsection{Chromatism}
Another downside is chromatism. Pair-wise probing usually works with narrowband filters, used subsequently, limiting the telescope duty cycle. For example, RST/CGI takes advantage of three $\sim18$~nm filters that are employed sequentially to perform wavefront sensing before applying a larger 56.5~nm (10\% bandwidth) imaging filter \cite{Cady2025}. In the SPHERE context, the dual-band imager IRDIS enables the use of the H2 filter in parallel to H3 to increase the bandwidth to 7.8\% at no cost. But, the current version of the PWP algorithm cannot directly work in high-bandwidth filters such as H-band filters (17.8\%) with IRDIS. Future work will thus rely on the SPHERE integral field spectrograph (IFS) to expand the working bandwidth up to 41\% \citep{Mesa2015}.

\subsubsection{Field of view}
The final downside is the limited obtained FoV. While adding a third probe in the algorithm enabled reaching a 360\degree post-processed image, the outer working angle of the corrected region is limited to the HODM influence function. This is acceptable since it corresponds to the AO correction region, where planets will be searched for with future XAO systems \citep{Boccaletti2020,Kasper2021}.

\section{Conclusion}
\label{sec:conclusion}
In this paper, we validated the CDI method relying on the temporal modulation of the VLT/SPHERE XAO HODM on astrophysical targets. Proposing a framework similar to those used in more classical post-processing algorithms (ADI and RDI), we combined a dozen sequences of reference images, each representing the speckle intensity distribution estimated by PWP and subtracted from raw images. By retrieving four almost face-on disks, we demonstrated that the post-processing technique is robust and does not suffer from self-subtraction, unlike commonly used ADI, while improving the image quality compared to noADI. Additionally, we discussed potential workarounds to increase the instrument duty cycle in future surveys with CDI and plans for future studies regarding the nonlinearity and limited bandwidth of the algorithm. We believe CDI is an efficient approach for direct imaging and advocate its use in the context of future facilities equipped with cutting-edge XAO instruments such as SAXO+ \citep{Boccaletti2020,Mazoyer2024}. It is worth noting that CDI should be even more efficient in stable conditions, such as those encountered in space-based instruments. Coherent differential imaging could then increase the duty cycle of such instruments, either by being applied while digging a DH directly on the target star \citep{Laginja2025}, and/or while stabilizing the DH \citep{Redmond2024}, for the coronagraph instrument on the Roman Space Telescope, or the Habitable Worlds Observatory.

\begin{acknowledgements}
The authors thank the anonymous referee for their valuable comments and suggestions, which have significantly improved this manuscript. This study was supported by the IdEx Université Paris Cité, ANR-18-IDEX-0001, and the Physics department of Université Paris Cité, and the Action Spécifique Haute Résolution Angulaire (ASHRA) of CNRS/INSU co-funded by CNES as well as an ECOS-CONICYT grant (\#C20U02). L.P., J.M., G.C. acknowledge support from Programa de Cooperación Científica ECOS-ANID ECOS200049. L.P.  acknowledges support by the ANID BASAL project FB210003 and ANID FONDECYT Regular 1221442. J. G. Kühn and R. Tandon have received funding from the Swiss State Secretariat for Education, Research, and Innovation as a SERI-Funded ERC 2021 Consolidator Grant, project RACE-GO \# M822.00084, following the discontinued participation of Switzerland to Horizon Europe.
\end{acknowledgements}

\bibliographystyle{aa} 
\bibliography{bib_AP} 

\appendix
\section{PWP algorithm}
\label{sec:PWP_algorithm}
Building on the idea developed by \cite{Borde2006} and first described by \cite{GiveOn2007SPIE}, the Pair-Wise probing (PWP) algorithm assumes low-amplitude aberrations and diversities relative to the wavelength. 
Let $\psi_m$ be the diversity introduced by a deformable mirror located in the pupil plane. The total intensity on the detector $I_m$ can then be written as
\begin{equation}
\label{eq:image_probe}
I_m = \left(E_S + iC[A\psi_m]\right)\left(E_S + iC[A\psi_m]\right)^* + I_a,
\end{equation}
where $E_S$ is the speckle E-field to be estimated, $C$ the linear operator that transforms a complex E-field in the pupil plane to a complex E-field in the focal plane through a coronagraph, and $A$ is the complex amplitude in the entrance pupil plane. $I_a$ represents the intensity of the circumstellar scene and the AO incoherent residuals. It fluctuates weakly with the applied probe and is incoherent with the stellar signal. If we record the pair of images $I_{m+}$ and $I_{m-}$, with respective diversities $+\psi_m$ and $-\psi_m$, the difference then becomes
\begin{equation}
I_{m+} - I_{m-} = 4 \left[ \Re(E_S) \Re(iC[A\psi_m]) + \Im(E_S) \Im(iC[A\psi_m]) \right],
\end{equation}
where $\Re$ and $\Im$ denote the real and imaginary parts of a complex number, respectively. We note that this image difference explicitly removes the astrophysical component $I_a$, ensuring no circumstellar scene in the final reconstructed coherent image.
This equation can be rewritten in matrix form:
\begin{equation}
I_{m+} - I_{m-} = 4 
\begin{bmatrix}
\Re(iC[A\psi_m]) & \Im(iC[A\psi_m])
\end{bmatrix}
\begin{bmatrix}
\Re(E_S) \\
\Im(E_S)
\end{bmatrix}.
\end{equation}
This system is undetermined and admits infinitely many solutions due to degeneracy between the real and imaginary parts of $E_S$.
Generalizing to $j$ pairs of images, for each pixel $(u,v)$ on the detector,
\begin{equation}
D_{(u,v)} = 4 M_{(u,v)} F_{(u,v)},
\end{equation}
where
\begin{equation}
D = \begin{bmatrix}
I_1^+ - I_1^- \\
\vdots \\
I_j^+ - I_j^-
\end{bmatrix},
\end{equation}
\begin{equation}
M = 
\begin{bmatrix}
\Re(iC[A\psi_1]) & \Im(iC[A\psi_1]) \\
\vdots & \vdots \\
\Re(iC[A\psi_j]) & \Im(iC[A\psi_j])
\end{bmatrix},
\end{equation}
\begin{equation}
F = 
\begin{bmatrix}
\Re(E_S) \\
\Im(E_S)
\end{bmatrix}.
\end{equation}
To reconstruct the E-field, we minimize the cost function:
\begin{equation}
J_{PW} = \min_F \left\| D - 4M \cdot F \right\|^2.
\end{equation}
This inverse problem is linear. It can be solved using the Singular Value Decomposition (SVD), provided that the matrix $M$ is invertible, i.e., its determinant is nonzero: 
\begin{equation}
\label{eq:PWP_matmul}
\tilde{F}_{(u,v)} = \frac{1}{4} M^\dagger_{(u,v)} D_{(u,v)}.
\end{equation}
Here, $\tilde{F}$ is the approximate reconstruction of $F$, and $M^\dagger$ is the Moore–Penrose pseudoinverse of $M$. The matrix $M$ is invertible at each pixel $(u,v)$, if there exists a pair of probes $(m,n)$ such that
\begin{multline}
\Re(iC[A\psi_m])_{(k,l)} \cdot \Im(iC[A\psi_n])_{(k,l)},\\
- \Re(iC[A\psi_n])_{(k,l)} \cdot \Im(iC[A\psi_m])_{(k,l)} \neq 0 .
\end{multline}
This condition means that at least two of the probes induce different E-fields at that location.

\section{Observing sequences}
\label{sec:observing_sequences}
\begin{figure}
    \centering
    \includegraphics[width=\linewidth]{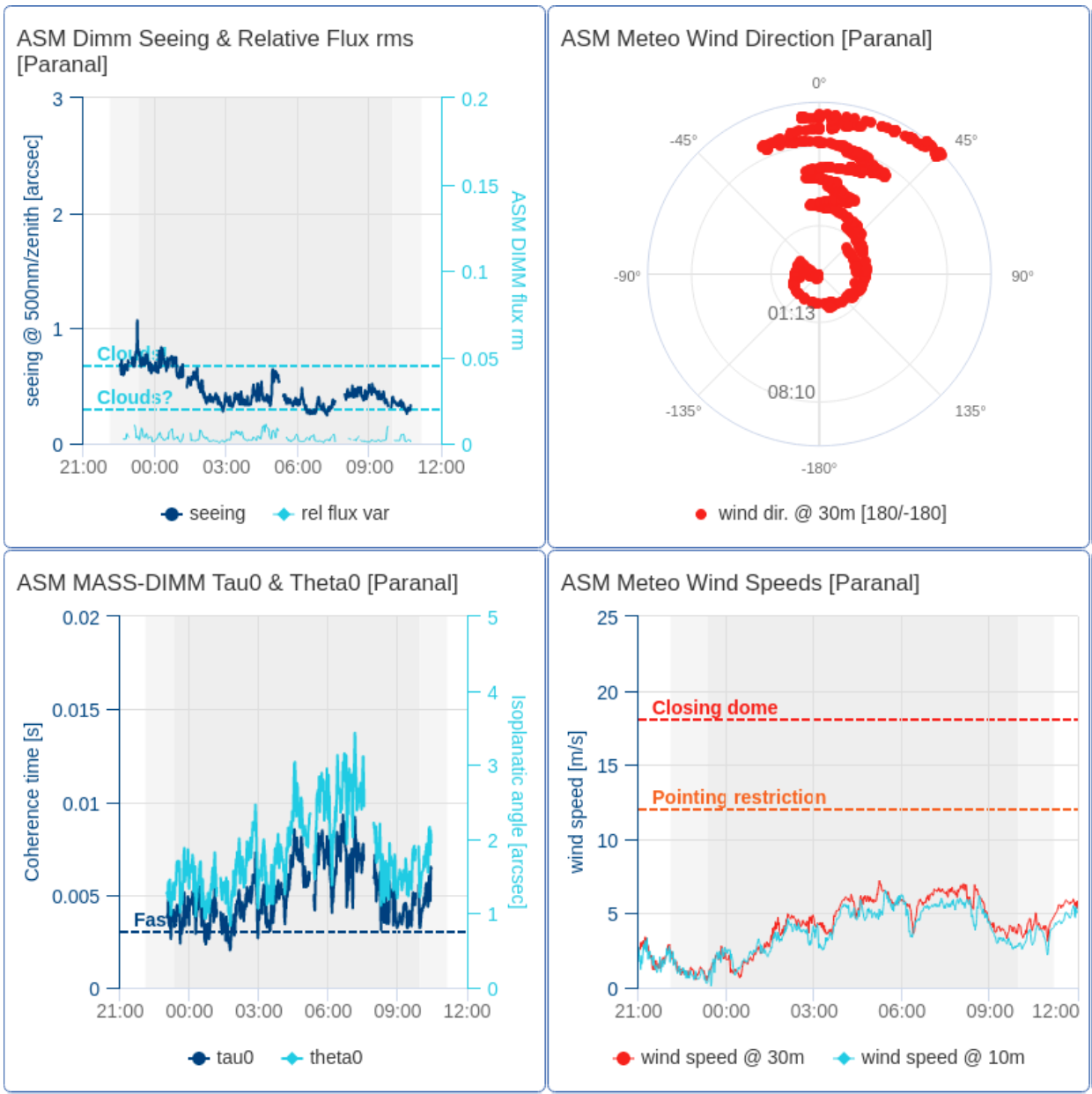}
    \caption{Observing conditions provided by the ESO UT3 visitor telescope report on June 4-5, 2024.}
    \label{fig:Observing_conditions}
\end{figure}

The observations were conducted during the night of June 4, 2024. According to the Paranal DIMM-Seeing monitor (see in Fig.~\ref{fig:Observing_conditions}), the atmospheric seeing ranged from 0.245'' to 0.834'' at 500~nm, improving rapidly to below 0.5'' after 00:00 UTC. As a result, the conditions were less favorable during the acquisition of HR~4796A data compared to those for HD~163296. These values reflect exceptional observing conditions at Paranal. The atmospheric coherence time increased steadily over the course of the night, ranging from approximately 5 ms to nearly 10 ms during the full observing sequence. Wind conditions were stable, originating from the north with speeds remaining below 5~m.s$^{-1}$. Observations were carried out using the H3 filter ($\lambda_0=1667$~nm, $\Delta\lambda=54$~nm) in combination with an Apodized Pupil Lyot Coronagraph \citep[APLC,][]{Soummer2005} in the APO1/ALC2 configuration, featuring a focal plane occulting mask with a diameter of 185 mas.

\section{Forward modeling of HR 4796A disk}
\begin{figure}
    \centering
    \includegraphics[width=\linewidth]{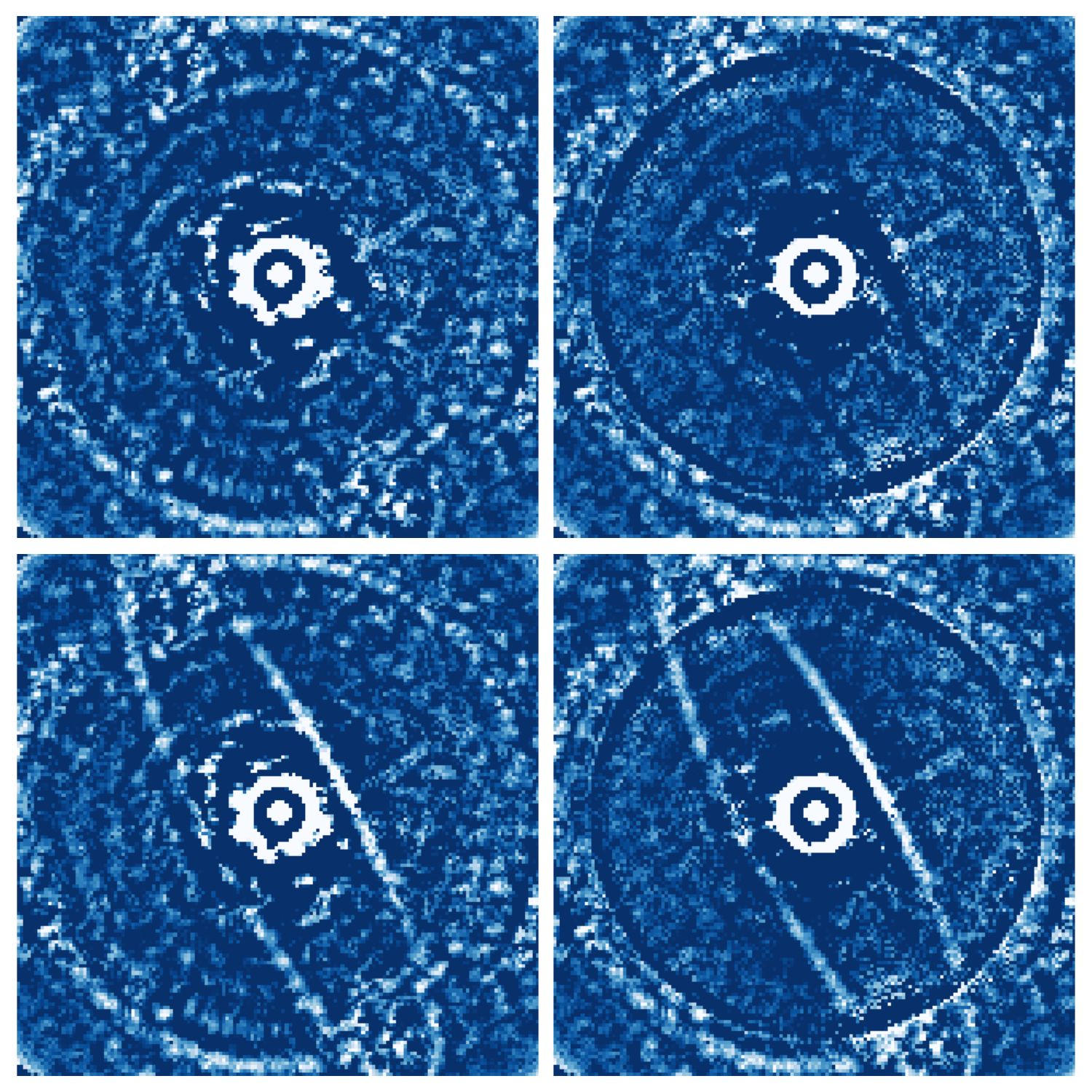}
    \caption{Forward modeling of HR 4796A. Shown are NoADI (left column) and CDI (right column) results before (bottom row) and after (top row) synthetic HR 4796A disk subtraction. We note a slight over-subtraction on the forward side.}
    \label{fig:FM_HR4796}
\end{figure}

Coherent-differential imaging performance gain is hard to assess in images with extended structures as obtained during our experiment. We therefore decided to produce a synthetic image cube of the disk around HR 4796 at specific position angles to simulate a real observation. This cube is then used to numerically subtract the disk signal from $I_{tot}$. We used the ray-tracing approach from the ScatteredLightDisk class in vip\_hci. Disks parameters from \cite{Milli2017} have been used to feed the function and have been slightly adjusted to minimize the disk signal by eye in the resulting image. Final images with and without CDI obtained through such process are shown in Fig.\ref{fig:FM_HR4796} before and after disk subtraction. The HR 4796A disk-free images are then used to quantify the throughput performance of the method.
\label{sec:FM_HR4796A}

\section{Framework for probe self-calibration}
\label{sec:selfcalibrationderivation}
Coherent-differential imaging self-calibration sequences can also be combined to increase the S/N , in a similar framework as in Sec.~\ref{subsubsec:PCA_coherent}. Here we assume three libraries containing the $N$ difference images for each probes $m$, whose singular value decomposition is given by

\begin{equation}
\label{eq:SVDdeltaI}
    \mathcal{M}_m = [\Delta I_{m,1}, \Delta I_{m,2}, ..., \Delta I_{m,N}] =  \mathcal{U}_mZ_m\mathcal{V}_m^T,
\end{equation}
where $\Delta I_{m,j}$ represents the $m^{\text{th}}$ probe image difference for the $j^{\text{th}}$ CDI sequence.  $\mathcal{U}_m$ and $\mathcal{V}_m$ are the left and right singular vectors, while $Z_m$ are the singular values in ascending orders. This decomposition can be used to filter out the noise components in the $\Delta I_m$, assuming noise to be represented in the ($N-K_{klip}$) last principle components of $\mathcal{M}_m $. Mathematically, $\Delta \mathcal{I}_{m,j}(K_{klip})$, the $m^{\text{th}}$ filtered probe image difference for the $j^{\text{th}}$ CDI sequence then becomes
\begin{equation}
    \Delta \mathcal{I}_{m,j}(K_{klip}) = \sum_{k=1}^{K_{klip}}<\Delta I_{m,j},\mathcal{V}^T_{m,k}>\mathcal{V}^T_{m,k} = (\left.\mathcal{U}_mZ_{m,K_{klip}}\mathcal{V}_m^T)\right|_{j}.
\end{equation}

As in Appendix~\ref{sec:PWP_algorithm}, the matrix of filtered probe image difference $\mathcal{D}_j(K_{klip})$ is computed via
\begin{equation}
\mathcal{D}_j(K_{klip}) = \begin{bmatrix}
\Delta\mathcal{I}_{1,j}(K_{klip}) \\
\vdots \\
\Delta\mathcal{I}_{m,j}(K_{klip})
\end{bmatrix},
\end{equation}
to perform a new E-field estimation $\mathcal{E}_j(K_{klip})$ at each sequence $j$ , leveraging the same $M$ probe model matrix used in usual PWP: 
\begin{equation}
\mathcal{E}_j(K_{klip})_{(u,v)} = \frac{1}{4} M^\dagger_{(u,v)} \mathcal{D}_j(K_{klip})_{(u,v)} .
\end{equation}
Such an E-field can now be added to the probe E-field model to calibrate for any probe images taken during the $N$ CDI sequences and retrieve the astrophysical signal $I_{a,mjl}$ from the probe image $I_{mjl}$:
\begin{equation}
    I_{a,mjl} = I_{mjl} - \left|\mathcal{E}_j(K_{klip})+(-1)^liC[A\psi_m]\right|^2,
\end{equation}
where $I_{mjl}$ is the $m^{th}$ probe of the $j^{th}$ CDI sequence, whose sign is defined with the argument $l$ ($l=1$ (resp. $l=2$) meaning negative (resp. positive) probe). We then chose to derotate each obtained solution while all $N$ sequences of six probe images are averaged out to form the astrophysical scene $I_a$ via
\begin{equation}
\label{eq:selfcalibrationderivation}
    I_a = \frac{1}{6N}\sum_{j=1}^N\sum_{m=1}^3\sum_{l=1}^2R_{\theta_j} I_{a,mjl}\quad.
\end{equation}

\end{document}